\documentclass[final,,twoside]{IEEEtranTCOM}
\normalsize
\ifCLASSINFOpdf
\else
\fi

\usepackage{amsthm}
\usepackage{amsmath}
\usepackage{amssymb}
\usepackage{graphicx}
\usepackage{esint}
\usepackage{amsfonts}
\usepackage{cite}
\usepackage{balance}
\usepackage{caption}
\usepackage{subcaption}
\usepackage{epstopdf}
\usepackage{color}
\makeatletter

\theoremstyle{plain}
\newtheorem{thm}{\protect\theoremname}

\makeatother

\providecommand{\theoremname}{Theorem}

\theoremstyle{plain}
\newtheorem{lem}{\protect\lemmaname}

\makeatother

\providecommand{\lemmaname}{Lemma}

\DeclareMathOperator{\card}{card}

\begin{document}
\title{Spectrum Sensing with Multiple Primary Users over Fading Channels}


\author{Alexandros~--~Apostolos~A.~Boulogeorgos,~\IEEEmembership{Student~Member,~IEEE,}~Nestor D. Chatzidiamantis, \IEEEmembership{Member, IEEE,} and George K. Karagiannidis, \IEEEmembership{Fellow, IEEE} 

\thanks{The authors are with the Department of Electrical and Computer Engineering, Aristotle University of Thessaloniki,  54 124, Thessaloniki, Greece (e-mail:{ \{ampoulog, nestoras, geokarag\}@auth.gr}).
}

}
\maketitle	

\begin{abstract}
We investigate the impact of multiple primary users (PUs) and fading on the spectrum sensing of a classical energy detector (ED). 
Specifically, we present novel closed-form expressions for the false-alarm and detection probabilities in a multiple PUs environment, assuming Nakagami-$m$ fading and complex Gaussian PUs transmitted signals. 
The results reveal the importance of taking into consideration the wireless environment,  when evaluating the ED spectrum sensing performance and selecting the ED~threshold.

\end{abstract}


\section{Introduction}\label{S:Intro}
\IEEEPARstart{T}{he} rapid growth of wireless communications and the foreseen spectrum occupancy problems, due to the exponentially increasing consumer demands on mobile traffic and data, inspired the evolution of the concept of cognitive radio (CR).
One fundamental task in CR that allows  the  exploitation  of  the  under-utilized  spectrum, is spectrum sensing. As a result, great amount of effort has been put to derive optimal, sub-optimal and ad-hoc solutions to the spectrum sensing problem and investigate their performance~\cite{Mutiuser_CR_networks_Joint_Impact_of_Direct_and_Relay_Communications,
A:Achievable_rates_and_OP_of_CR,
A:CRNs_withs_Multi_PUs_under_Spectrum_sharing,
A:Relay_selection_for_Sec_in_CRNs,
A:Outage_Perf_of_Spect_Efficient_Schemes_for_Multiuser_CRNs,
A:Performance_of_User_Selection_in_Cognitive_Broadcast_Channels}.

Scanning the open literature, most of the related works have neglected the impact of multiple primary users (PUs) and fading on the spectrum sensing performance of the CR's energy detector (ED). 
However, in several widely used wireless communication standards, such as long term evolution advanced (LTE-A), WiFi and WiMAX, where code-division-multiple-access (CDMA) is used, users simultaneously operate in the same frequency band. 
These applications motivated a general investigation of the effect of PU traffic on the sensing performance, when multiple PUs are present.
To the best of the authors' knowledge, there is only one published work in the open literature \cite{A:Multiple_PUs_Spectrum_Sensing}, where the  effect of multiple PUs on spectrum-sensing performance was investigated, considering only the impact of additive white Gaussian  noise  (AWGN)  channels. Moreover, in \cite{Cacciapuoti2015}, the authors derived the sensing time and the transmission time that jointly maximize
the sensing efficiency and the sensing accuracy in a multiple mobile PU network. However, in \cite{Cacciapuoti2015}, the spectrum sensing method and the effect of fading channels was neglected.

In this letter, we present an analytical framework to evaluate and quantify the effects of multiple PUs and fading on the spectrum sensing performance of a classical ED.
In particular, we present novel closed-form expressions for the false-alarm and detection probabilities in a multiple PUs environment, assuming Nakagami-$m$ channels and complex Gaussian transmitted signals. 

\subsubsection*{Notations} 
Unless otherwise stated, $\Re\left\{ x\right\} $ and $\Im\left\{ x\right\} $ represent the real and imaginary part of $x$,  
operators $E\left[\cdot\right]$ and $\left|\cdot\right|$ denote the statistical expectation and the absolute value, respectively, while the operator $\exp\left(\cdot\right)$ denotes the exponential function.
The operator $\card\left(\mathcal{A}\right)$ returns the  cardinality of  the set $\mathcal{A}$.
The lower \cite[Eq. (8.350/1)]{B:Gra_Ryz_Book} and upper incomplete Gamma functions \cite[Eq. (8.350/2)]{B:Gra_Ryz_Book} are represented by $\gamma\left(\cdot,\cdot\right)$ and $\Gamma\left(\cdot,\cdot\right)$, respectively, while the Gamma function \cite[Eq. (8.310)]{B:Gra_Ryz_Book} is denoted by $\Gamma\left(\cdot\right)$. 
Finally, $\Gamma\left(\cdot,\cdot,\cdot,\cdot\right)$ is the extended incomplete Gamma function defined in \cite[Eq. (6.2)]{B:chaudhry2001class}.

\section{System and signal model}\label{sec:SSM}

We consider a multiple PUs/secondary user (SU) environment, where $M$ static PUs operate in the same frequency band, which is sensed by a single CR device.  
The two possible states, i.e., busy or idle, of the $i-$th PU are denoted with the parameters $\theta_i\in\left\{0,1\right\}$, where $i=1,2,\ldots,M$.  Suppose that the $n$-th sample of the transmitted signal of the $i-$th PU, $s_i\left(n\right),$ is conveyed over a flat-fading wireless channel, with channel gain, $h_i\left(n\right)$. Hence, at the SU detector the $n-$th sample of the baseband equivalent received signal can be expressed~as
\begin{equation}
y(n) = \sum_{i=1}^{M}\theta_i h_i(n) d_i^{-\xi_i/2} s_i(n) + w(n),
\label{Eq:Received_signal_multiple_PUs}
\end{equation}
where $d_i$ and $\xi_i$ stand for the distance between the $i-$th PU and the SU, and the corresponding link path-loss exponent, respectively, while $w(n)$ represents the AWGN. 
We assume that $s_i$ and $w$ are zero-mean circular symmetric complex white Gaussian processes with variances $\sigma_{s_i}^2$ and $\sigma_w^2$. Furthermore, $h_i$ is a zero mean complex random variable (RV) with variance $\sigma_{h_i}^2$ and $|h_i|$ follows Nakagami-$m_i$~distribution. 
Without loss of generality, it is assumed that  SU is located in the first ($i=1$) PU cell. 

Next, $\Theta=\left[\theta_1, \theta_2, \cdots, \theta_M\right]$ represents the set of $M$ PUs (busy and idle) located at distances $d=\left[d_1, d_2, \cdots, d_M\right]$ from the SU, 
while $\tilde{\Theta}=\left[\tilde{\theta}_1, \tilde{\theta}_2, \cdots, \tilde{\theta}_L\right]\subseteq\Theta$ denotes the set of the $L\leq M$ active PU located at distances $\tilde{d}=\left[\tilde{d}_1, \tilde{d}_2, \cdots, \tilde{d}_L\right]$.
Additionally, $\Theta_0=\left[0, 0, \cdots, 0\right]$ stands for the all idle PU occupancy set. $\Theta_1=\left[1,\theta_2, \cdots, \theta_M\right]$, with $\theta_j\in\{0,1\}$, $j=2,\cdots,M$, represents the PU occupancy set, in which the first PU is active, while $\Theta_{0,1}=\left[0,\theta_2, \cdots, \theta_M\right]$, in which at least one $\theta_l=1$, $l\in\{2, \cdots, M\}$, denotes the PU occupancy set in which the first PU is idle and at least one PU is busy. 
Finally, $\tilde{\Theta}_1$ and $\tilde{\Theta}_{0,1}$ denote the corresponding to $\Theta_{1}$ and $\Theta_{0,1}$ sets of active~PUs.

\section{False Alarm/Detection Probabilities\label{sec:Probabilities}}

In the classical ED, the energy of the received signals is used to determine whether a channel is idle or busy. Based on the signal model described in Section \ref{sec:SSM}, ED calculates the energy test statistics~as
\begin{align}
T & \hspace{-0.12cm}=\hspace{-0.12cm}\frac{1}{N_{s}}\hspace{-0.12cm}\sum_{n=0}^{N_{s}-1}\left|y\left(n\right)\right|^{2}
\hspace{-0.12cm}=\hspace{-0.12cm}\frac{1}{N_{s}}\hspace{-0.12cm}\sum_{n=0}^{N_{s}-1}\Re\left\{ y\left(n\right)\right\} ^{2}+\Im\left\{ y\left(n\right)\right\} ^{2},\label{ED_classic}
\end{align}
where $N_{s}$ is the number of samples used for spectrum sensing. 
The energy test statistic, $T$, is compared against a threshold $\gamma$ to yield the sensing decision, 
i.e., the ED decides that the channel is busy if $T>\gamma$ or idle,~otherwise.

For a given channel realization set $H=\left\{ h_{1},h_{2},\cdots,h_{M} \right\}$ and PUs occupation set $\Theta=\left\{\theta_1,\theta_2,\cdots,\theta_M\right\}$, the real and imaginary parts of the received signals are uncorrelated, i.e., $E\left[\Re\left\{ y\right\} \Im\left\{ y\right\} \right]=0$, with variances
\begin{align}
E\left[\Re\left\{ y\right\} ^{2}\right]&=E\left[\Im\left\{ y\right\} ^{2}\right]=\sigma^{2} 
\nonumber \\& 
=\sum_{i=1}^{M}\theta_i\left|h_i\right|^2 d_i^{-\xi_i} \frac{\sigma_{s_i}^2}{2}+\frac{\sigma_{w}^{2}}{2},
\label{Eq:sigma_MPU}
\end{align}
the received energy test statistic follows chi-square distribution with $2N_{s}$ degrees of freedom and cumulative distribution function (CDF) given by 
\begin{align}
&F_{T}\left(x\left|H,\Theta\right.\right)
=\frac{\gamma\left(N_{s},\frac{N_{s}x}{2\sigma^{2}}\right)}{\Gamma\left(N_{s}\right)}.
\label{eq:ideal_cond0}
\end{align}
Furthermore, since $N_s$ is an integer, \eqref{eq:ideal_cond0} can be re-written as~\cite[Eq. (8.352/2)]{B:Gra_Ryz_Book}
\begin{align}
F_{T}\left(x\left|H,\Theta\right.\right)=1-\sum_{n=0}^{N_s-1}&\frac{1}{n!}\left(\frac{N_s x}{\sigma^2}\right)^n 
\exp\left(-\frac{N_s x}{\sigma^2}\right).\label{eq:ideal_cond}
\end{align}


\begin{thm}
\label{thm:CDF_MPU} The CDF of the energy test statistics for a given channel set, $\tilde{\Theta}\subseteq\Theta$, with $L\in[1,M]$ active PUs, can be evaluated by \eqref{CDF_theta_M_MU}, given at the top of the next page,
\begin{figure*}
\begin{align}
&F_{T}\left(x\left|\Theta\right.\right) = 1 - \sum_{n=0}^{N_s-1} \sum_{i=1}^{L}\sum_{k=1}^{a_{i}} \sum_{j=0}^{k-1}\frac{\left(-c\right)^{k-1-j} \Xi(i,k)}{n! b_i^{k+n-j-1} (k-1)!}\left(\begin{array}{c}k-1\\j\end{array}\right) 
\left(\frac{N_s x}{2}\right)^n   \exp\left(\frac{c}{b_i}\right) 
\Gamma\left(-n+j+1,\frac{c}{b_i},\frac{N_s x}{2 b_i},1\right)
\label{CDF_theta_M_MU}
\end{align}
\vspace{-0.1cm}
\hrulefill
\vspace{-0.4cm}
\end{figure*}
where $a=\{\tilde{m}_1, \tilde{m}_2,\cdots, \tilde{m}_{L}\}$, 
$b=\left\{ \frac{ \tilde{d}_1^{-\tilde{\xi}_1} \sigma_{\tilde{h}_1}^2 \sigma_{\tilde{s}_1}^2}{2 \tilde{m}_1}, \frac{\tilde{d}_2^{-\tilde{\xi}_2} \sigma_{\tilde{h}_2}^2 \sigma_{\tilde{s}_2}^2}{2 \tilde{m}_2}, \cdots, \frac{\tilde{d}_L^{-\tilde{\xi}_L} \sigma_{\tilde{h}_L}^2 \sigma_{\tilde{s}_L}^2}{2 \tilde{m}_L} \right\}$, and $c=\frac{\sigma_w^2}{2}$. Furthermore, $\tilde{d}_i$ and $\tilde{\xi}_i$, represent the distance and the corresponding link path-loss exponent between the $i-$th active PU and the ED, whereas  $\sigma_{\tilde{s}_i}^2$, and $\sigma_{\tilde{h}_i}^2$ stand for the variances of the $i-$th active PU's transmitted signal and $i-$th fading channel.
The shape factor of the fading channel between the $i-$th active PU and the CR device is denoted as $\tilde{m}_i$.
Moreover, note that in \eqref{CDF_theta_M_MU}, $\Xi(i,k)$ is defined in~\cite[Eqs. (8) and (9)]{A:Karagiannidis-2006-ID448}
\footnote{Note that there is a typo in \cite[Eq. (8)]{A:Karagiannidis-2006-ID448}. The correct expression is
\begin{align*}
\Xi(i,a_i - k) \hspace{-0.05cm}=\hspace{-0.05cm} \frac{1}{k}\hspace{-0.1cm} \sum_{\begin{array}{c}q=1\\q\neq i\end{array}}^{L} \sum_{j=1}^{k} \frac{a_q}{b_{i}^j} 
\left(\frac{1}{b_i} - \frac{1}{b_q}\right)^{-j} \Xi(i,a_i-k+j). 
\end{align*}}.
\end{thm}

\begin{IEEEproof}
Since $|h_i|$ is a zero mean Nakagami-$m$ distributed RV, the variance of the received signal, given by \eqref{Eq:sigma_MPU}, is a sum of squared Nakagami-$m$ RV with probability density function (PDF) given~by \cite{A:Karagiannidis-2006-ID448}
\begin{align}
f_{\sigma^2}(y) = \sum_{i=1}^{L}\sum_{k=1}^{a_{i}} \Xi\left(i,k\right) \frac{\left(y-c\right)^{k-1}}{b_i^k\left(k-1\right)!}\exp\left(-\frac{y-c}{b_i}\right), 
\label{Eq:PDF_generalized}
\end{align}
with $y\in\left[c,\infty\right)$. Hence, the unconditional CDF of the energy test statistic, $T$, can be expressed as 
\begin{align}
F_{T}\left(x\left|\Theta\right.\right) &= 
\int_{c}^{\infty}f_{\sigma^{2}}\left(y\right)dy 
- \sum_{n=0}^{N_s-1}\frac{1}{n!}\left(\frac{N_s x}{2}\right)^n 
\nonumber \\ & \times
\int_{c}^{\infty} y^{-n} \exp\left(-\frac{N_s x}{2 y}\right)f_{\sigma^{2}}\left(y\right)dy.
\label{Eq:CDF_generalized_step_1}
\end{align}
Since $y\in[c,\infty)$, $\int_{c}^{\infty}f_{\sigma^{2}}\left(y\right)dy = 1$. Additionally, by substituting \eqref{Eq:PDF_generalized} into \eqref{Eq:CDF_generalized_step_1} and after some mathematical manipulations, \eqref{Eq:CDF_generalized_step_1} yields
\begin{align}
F_{T}\left(x\left|\Theta\right.\right) &=  1 - \sum_{n=0}^{N_s-1} \sum_{i=1}^{L}\sum_{k=1}^{a_{i}}  \frac{\Xi\left(i,k\right)}{b_i^k\left(k-1\right)! n!}\left(\frac{N_s x}{2}\right)^n 
\nonumber \\ &\hspace{-1cm}\times
\int_{c}^{\infty} y^{-n} \left(y-c\right)^{k-1} \exp\left(-\frac{y-c}{b_i}-\frac{N_s x}{2 y}\right)dy. 
\label{Eq:CDF_theta_1_MU_s1}
\end{align}
Since $k$ is an integer and $k\geq 1$, by using the binomial expansion in $\left(y-c\right)^{k-1}$, \eqref{Eq:CDF_theta_1_MU_s1} can equivalently be written as
\begin{align}
F_{T}\left(x\left|\Theta\right.\right) = 
&
1 - \sum_{n=0}^{N_s-1} \sum_{i=1}^{L}\sum_{k=1}^{a_{i}} \sum_{j=0}^{k-1}\frac{\Xi(i,k)}{n! b_i^k (k-1)!}\left(\begin{array}{c}k-1\\j\end{array}\right) 
\nonumber \\ & \times
\left(\frac{N_s x}{2}\right)^n \left(-c\right)^{k-1-j} \exp\left(\frac{c}{b_i}\right)
\nonumber \\ & \times
\int_{c}^{\infty}y^{-n+j}\exp\left(-\frac{N_s x}{2 y} - \frac{y}{b_i}\right) dy.
\label{Eq:CDF_theta_1_MU_s2}
\end{align}
Finally, by setting $z=\frac{y}{b_{i}}$ into \eqref{Eq:CDF_theta_1_MU_s2} and using \cite[Eq. (6.2)]{B:chaudhry2001class}, \eqref{Eq:CDF_theta_1_MU_s2} yields \eqref{CDF_theta_M_MU}.
This concludes the proof.
\end{IEEEproof}
Note that  $\Gamma\left(\cdot,\cdot,\cdot,1\right)$ can be evaluated by~\cite[Theorem.4]{DBLP:journals/corr/BoulogeorgosCK15}.

\begin{lem}
The CDF of the energy test statistic assuming all the PUs are idle can be evaluated by
\begin{align}
F_{T}\left(x\left|\Theta_0\right.\right) = 1-\sum_{n=0}^{N_s-1}\frac{1}{n!}\left(\frac{N_s x}{\sigma_w^2}\right)^n \exp\left(-\frac{N_s x}{\sigma_w^2}\right).
\label{Eq:CDF_FA_MPU_all_idle}
\end{align}
\end{lem}
\begin{IEEEproof}
If $\Theta=\Theta_0$, according to \eqref{Eq:sigma_MPU}, $\sigma^2=\frac{\sigma_w^2}{2}$, which is independent of the channel realization set $H$. Substituting this value into \eqref{eq:ideal_cond}, we get \eqref{Eq:CDF_FA_MPU_all_idle}. This concludes the proof.
\end{IEEEproof}

Based on the above analysis the detection and false-alarm probabilities can be respectively obtained by
\begin{align}
P_{\mathbf{d}}(\gamma) = \sum_{i=1}^{\card\left(\tilde{\Theta}_1\right)}P_{r}\left({\Theta_1}\right) 
\left(1-F_{T}\left(\gamma \left| \tilde{\Theta}_1\right. \right) \right)
\label{Eq:Detection_Probability_M_PU}
\end{align}
and
\begin{align}
& P_{\mathbf{fa}}(\gamma) = \sum_{i=1}^{\card\left(\tilde{\Theta}_{0,1}\right)}P_{r}\left({\Theta}_{0,1}\right) 
\left(1-F_{T}\left(\gamma \left| \tilde{\Theta}_{0,1}\right. \right) \right) 
\nonumber \\ &
+ P_{r}\left(\Theta_0\right) \sum_{n=0}^{N_s-1}\frac{1}{n!}\left(\frac{N_s x}{\sigma_w^2}\right)^n \exp\left(-\frac{N_s x}{\sigma_w^2}\right),
\label{Eq:False_Alarm_Probability_M_PU}
\end{align}
where $P_{r}\left(\Theta\right)$ stands for the probability of the PU occupancy set $\Theta$, and $\tilde{\Theta}$ denote the set of active PUs. 
According to \eqref{Eq:Detection_Probability_M_PU}, \eqref{Eq:False_Alarm_Probability_M_PU} and \eqref{CDF_theta_M_MU}, the detection and false alarm probabilities depend not only take on the PU that is located in the same cell as the SU, but also the interference and the fading characteristics of the neighbor PUs-SU links.
Consequently, in order to select the energy detection threshold and the number of samples that will be used to achieve a detection and/or false alarm probabilities requirements, ED should take into consideration not only the variances of the PU signal, noise and channel, but also the variances of the neighbor PUs' signals,  and channels, as well as the probabilities of active PU existence. Note that in practice, the CR device has certain noise measurements and has only an estimate for the noise variance. 
However, in our analysis, we assume that the ED has perfect knowledge on the noise variance, which is obtained from calibration measurements. 
This is a typical assumption (\cite{DBLP:journals/corr/BoulogeorgosCK15} and references therein) in order to be able to quantify the performance degradation due to the effects of multiple PUs, independently of the classical noise uncertainty problem.
Next, we study two important special cases.

\underline{Special Case 1 (Rayleigh fading):} 
In the special case in which all the PUs-SU links are Rayleigh distributed, the CDF of the energy test statistics for the given set $\tilde{\Theta}$ can be obtained by setting $a=\{1,1, \cdots, 1\}$ into  \eqref{CDF_theta_M_MU}~as  
\begin{align}
F_{T}\left(x\left|\Theta\right.\right) = 1 - \sum_{n=0}^{N_s-1} & \sum_{i=1}^{L} 
\frac{ \Xi(i,1)}{n! b_i^{n}}
\left(\frac{N_s x}{2}\right)^n   \exp\left(\frac{c}{b_i}\right) 
\nonumber \\ & \times
\Gamma\left(-n+1,\frac{c}{b_i},\frac{N_s x}{2 b_i},1\right).
\label{CDF_theta_M_MU_Rayleigh}
\end{align}

\underline{Special Case 2 (single PU scenario):} In the special case of a single PU, the CDF of the energy test statistic assuming that the PU is active can be obtained by setting $L=1$, $a=\{m\}$ and $b=\left\{\frac{d^{-\xi \sigma_h^2 \sigma_s^2}}{2 m}\right\}$ into \eqref{CDF_theta_M_MU}, as \eqref{CDF_theta_1_ideal}, given at the top of the next page.

\begin{figure*}
\begin{align}
  F_{T}&\left(x\left|\theta\right.\right) = 1 - \frac{2 m^m}{\left(\sigma_h^2 d^{-\xi} \sigma_s^2\right)^{m} \Gamma(m)} \exp\left(\frac{m \sigma_w^2}{\sigma_h^2 d^{-\xi} \sigma_s^2}\right) 
 \nonumber \\ & \times
 \sum_{k=0}^{m-1}\sum_{n=0}^{N_s-1}
 \left(\begin{array}{c} m-1 \\ k\end{array}\right) \frac{2^{k-n}}{n!}\left(-\sigma_w^2\right)^{m-1-k} \left(N_s x\right)^{n} \left(\frac{\sigma_h^2 d^{-\xi} \sigma_s^2}{2m} \right)^{k-n+1} \Gamma\left(k-n+1,\frac{m\sigma_w^2}{\sigma_h^2 d^{-\xi} \sigma_s^2},\frac{m N_s x}{\sigma_h^2 d^{-\xi} \sigma_s^2},1\right)
\label{CDF_theta_1_ideal}
\end{align}
\vspace{-0.1cm}
\hrule
\vspace{-0.4cm}
\end{figure*}

Furthermore, the CDF of the energy test statistics assuming that the PU is idle can be derived by \eqref{Eq:CDF_FA_MPU_all_idle}. Therefore, the detection and false-alarm probabilities in the single PU scenario can be respectively obtained~as
\begin{align}
&P_{\mathbf{d}}(\gamma)  = \hspace{-0.1cm} P_r\left(T_{k}\hspace{-0.1cm} >\hspace{-0.1cm} \gamma\left|\theta_{k}\hspace{-0.1cm}=\hspace{-0.1cm}1 \right.\right)
\nonumber \\ & = 
\frac{2 m^m}{\left(\sigma_h^2 d^{-\xi} \sigma_s^2\right)^{m} \Gamma(m)} \exp\left(\frac{m \sigma_w^2}{\sigma_h^2 d^{-\xi} \sigma_s^2}\right) 
 \nonumber \\ & \times
 \sum_{k=0}^{m-1}\sum_{n=0}^{N_s-1}
 \left(\begin{array}{c} m-1 \\ k\end{array}\right) \frac{2^{k-n}}{n!}\left(-\sigma_w^2\right)^{m-1-k} \left(N_s \gamma\right)^{n} 
 \nonumber \\ & \times 
 \left(\frac{\sigma_h^2 d^{-\xi} \sigma_s^2}{2m} \right)^{k-n+1} \Gamma\left(k-n+1,\frac{m\sigma_w^2}{\sigma_h^2 d^{-\xi} \sigma_s^2},\frac{m N_s \gamma}{\sigma_h^2 d^{-\xi} \sigma_s^2},1\right),
 \label{Eq:P_d_Ideal_RF}
\end{align}
and 
\begin{align}
P_{\mathbf{fa}}(\gamma) & = \sum_{n=0}^{N_s-1}\frac{1}{n!}\left(\frac{N_s \gamma}{\sigma_w^2}\right)^n \exp\left(-\frac{N_s \gamma}{\sigma_w^2}\right).
\label{Eq:P_fa_Ideal_RF}
\end{align}
In order to meet the requirements for the detection and/or false alarm probabilities, the ED should appropriately set the detection threshold and the number of samples, 
by taking into consideration parameters as the PU signal variance, the noise variance, the fading characteristics and the path-loss exponent of the PU-SU link, as well as the distance between the PU and the~SU. 

Note that the single PU scenario has been extensively studied in the open literature, considering deterministic PU transmission signal. 
However, to the best of the authors' knowledge, this is the first work, in which a closed-form expression for the CDF of the energy statistics, under the assumptions of Nakagami-$m$ fading and complex Gaussian distributed PU transmitted signal, is presented.
Therefore, the derived expressions can be used to quantify the effects of Nakagami-$m$ fading, by neglecting the impact of multiple PUs.  
Finally, in Section \ref{sec:Numerical_Results}, the single PU scenario, is used as a benchmark to demonstrate the impact of multiple PUs on the spectrum sensing performance of the ED.
  
\section{Numerical and Simulation Results}\label{sec:Numerical_Results}

\begin{figure}
\centering\includegraphics[width=0.57\linewidth,trim=0 0 0 0,clip=false]{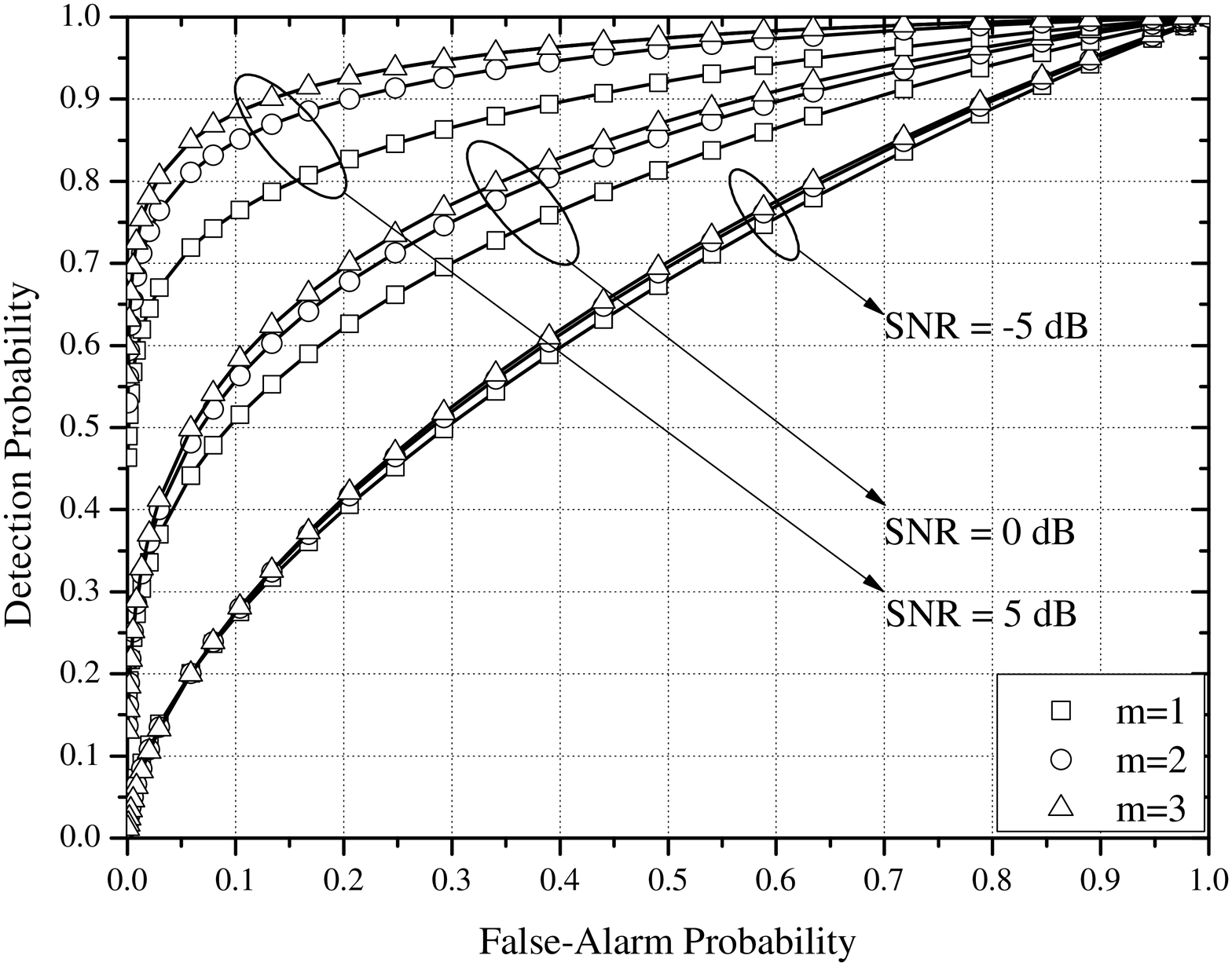}
\caption{ROCs for systems with a single PU and different values of $m$ and $\rm{SNR}$.}
\label{fig:ROC_1PU}
\end{figure}
In  this  section, using the previous results, we  investigate  the  impact of fading and the existence of multiple PUs on the spectrum sensing performance of EDs. 
For all figures, the number of samples is set to 5 ($N_s= 5$), while it is assumed that $\sigma^2_{h_i}=\sigma^2_w= 1$, $i=1,\ldots,M$. In all the illustrations, the solid curves represent analytical values obtained through the derived formulas, while the markers represent Monte-Carlo simulation results.

In Fig. \ref{fig:ROC_1PU}, receiver operation curves (ROCs) are demonstrated for different signal-to-noise ratios ($\rm{SNR}$s) and $m$ values, in the presence of a single PU, i.e., $M=1$. 
We observe that for low $\rm{SNR}$ values, the characteristics of the fading channels do not significantly affect the ED performance. 
However, as $\rm{SNR}$ increases, the effects of the fading statistics become more detrimental. In addition, it is seen that for a fixed $m$ and false-alarm probability, the detection probability of the ED increases as the $\rm{SNR}$ increases. 
Moreover, for a fixed $\rm{SNR}$ and false-alarm probability, as $m$ increases, the effects of fading become less severe; hence, the detection probability~increases.

\begin{figure}
\centering\includegraphics[width=0.57\linewidth,trim=0 0 0 0,clip=false]{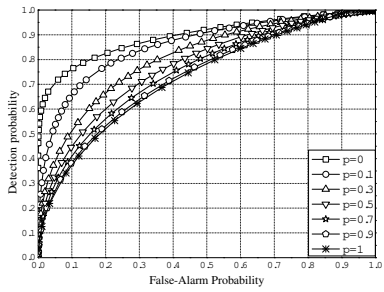}
\caption{ROCs for systems with $6$ PUs and different values of~$p$.}
\label{fig:ROC_6PU}
\end{figure}

Next, we consider that the CR device operates in a 6-PUs environment, where all the fading channels are assumed to be Rayleigh distributed ($m_i=1$, $i\in\{1,\ldots,6\}$) and each PU causes a different level of interference to the CR. We assume that the SU belongs to the first PU's cell, while the other $5$ PUs are interferers from neighbor cells. 
In particular, in Fig. \ref{fig:ROC_6PU}, ROCs are  plotted  for  different probabilities of active interfering PU existence, $p$, considering that the received $\rm{SNR}$ from the PU, which is located in the same cell with the SU, is equal to $0\text{ }\rm{dB}$, while the interference-to-noise ratios ($\rm{INR}$s) from the other $5$ PUs are $0\text{ }\rm{dB}$, $-1\text{ }\rm{dB}$, $-2\text{ }\rm{dB}$, $-3\text{ }\rm{dB}$ and $-5\text{ }\rm{dB}$. 
Note that we assume the same $p$ for the $5$ interfering PUs.
It is observed that as $p$ increases, the probability of interference of an neighbor PU increases; consequently, the spectrum sensing capabilities of the ED are constrained. 
For example, for a fixed $\rm{P}_{\rm{fa}}=0.1$, the detection probability is decreased about $41.8\%$ for $p=0.5$ in comparison with the case in which $p=0$. Notice that the $p=0$ case corresponds to the single PU scenario. 

\begin{figure}
\centering\includegraphics[width=0.57\linewidth,trim=0 0 0 0,clip=false]{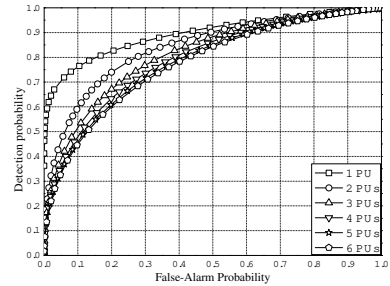}
\caption{ROCs for systems with $N$ PUs with $p=0.5$.}
\label{fig:ROC_NPU}
\end{figure}

In Fig. \ref{fig:ROC_NPU}, ROCs are illustrated for different number of PUs, $M$, considering that 
$m_i=1$, for $i=1,\ldots,6$, and the probability of existence of the $j-$th PU, $j\in\{2, \ldots, M\}$ is equal to 0.5, i.e., $p=0.5$. We observe that as the number of PUs increases, the interference from neighbor PUs increases; hence the false-alarm probability increases and the spectrum sensing capabilities of the ED are constrained.

\section{Conclusions}\label{sec:Conclusions}

In this letter, we studied the impact of multiple PUs in the spectrum sensing performance of a classical ED, assuming Nakagami-$m$ channels and complex Gaussian PUs' transmitted signals. 
Our results revealed the importance of taking into consideration the fading statistics, especially in the medium to high $\rm{SNR}$ regime. 
Furthermore, we observed that the spectrum sensing performance is constrained as the  probability of interference from neighbor PUs increases. 
Therefore, when selecting the operational energy detection threshold, we should not only take into consideration the PU that is located in the same cell as the SU, but also the wireless  environment, i.e.,  interference, as well as the fading~characteristics of the PUs-SU~links.

\vspace{-0.4cm}
\bibliographystyle{IEEEtran}
\bibliography{IEEEabrv,References}
\balance

\end{document}